\documentclass[preprint,showpacs,prl]{revtex4}
\usepackage{stmaryrd}
\usepackage{amsmath}
\usepackage{amssymb}
\usepackage{graphicx}
\usepackage{dcolumn}
\usepackage{bm}

\begin{document}

\begin{titlepage}

\title{Electronic Strengthening of Graphene by Charge Doping}

\author{Chen Si,$^{1,2}$ Wenhui Duan,$^1$ Zheng Liu,$^2$ and Feng Liu$^2$\footnote{fliu@eng.utah.edu}}
\address{$^1$Department of Physics and State Key Laboratory of Low-Dimensional Quantum Physics, Tsinghua University, Beijing 100084, People's Republic of China \\
$^2$Department of Materials Science and Engineering, University of Utah, Salt Lake City, Utah 84112, USA}

\date{\today}

\begin{abstract}
Graphene is known as the strongest 2D material in nature, yet we show that moderate charge doping of either electrons or holes can further enhance its ideal strength by up to $\sim$17\%, based on first principles calculations. This unusual electronic enhancement, versus conventional structural enhancement, of material's strength is achieved by an intriguing physical mechanism of charge doping counteracting on strain induced enhancement of Kohn anomaly, which leads to an overall stiffening of zone boundary K$_{1}$ phonon mode whose softening under strain is responsible for graphene failure. Electrons and holes work in the same way due to the high electron-hole symmetry around the Dirac point of graphene, while over doping may weaken the graphene by softening other phonon modes. Our findings uncover another fascinating property of graphene with broad implications in graphene-based electromechanical devices.
\end{abstract}



\maketitle

\draft

\vspace{2mm}

\end{titlepage}
Graphene is a single layer of sp$^{2}$-hybridized carbon atoms arranged in honeycomb lattice, with in-plane $\sigma$ bonds forming the skeletons of honeycomb and out-of-plane $\pi$ bonds forming conjugated 2D electron gas. This unique structure results in various fascinating properties of graphene\cite{NMT,RMP}. In particular, mechanically graphene possesses extremely high stiffness and strength and is known as the strongest 2D material in nature\cite{Lee}, which inspires a range of potential applications such as light-weight high-strength materials and composites. Recent theory reveals that the failure mechanism of graphene under tension lies in the softening instability of the zone-boundary K$_{1}$ phonon mode occurring at a critical strain of $\sim$15\%\cite{Marianetti}.

Search and design of hard materials has long attracted great interest with significant technological implications\cite{A.Y.Liu, Teter}. The ideal strength of materials is usually enhanced by structural designs at atomic and molecular level. For example, the carbon clathrate (C-46) is found to have a larger ideal strength than the diamond because its special cage structure effectively inhibits the instability of carbon sp$^{3}$ to sp$^{2}$ transition\cite{Blase}. Common strategies of material strengthening are to form composites\cite{Jones} or by alloying and doping, such as high-strength steels formed with different dopants in Fe-C\cite{Bain}. Here, using first-principle calculations, we demonstrate an unusual electronic enhancement of material's ideal strength by pure charge doping, in contrast to the conventional structural enhancement. We show that moderate charge doping ($\leq$ 10$^{14}$ cm$^{-2}$) of either electrons or holes can increase the strength of graphene by up to $\sim$17\%. We further show that the strain induced softening of zone-boundary K$_{1}$ mode, which was identified responsible for graphene failure\cite{Marianetti}, is associated with the strain enhanced Kohn anomaly in graphene. The surprising charge strengthening of graphene we discover is caused by doping induced suppression of Kohn anomaly, which counteracts the strain effect, and hence hardens the K$_{1}$ mode. Electrons and holes work in the same way due to the high electron-hole symmetry around the Dirac point of graphene.

Graphene is often doped either intentionally or unintentionally. When graphene is placed on a substrate\cite{Martin, Si} or adsorbed with foreign atoms and molecules\cite{Gierz, Leenaerts}, it is naturally doped due to charge transfer. In electronic devices, graphene is purposely doped by charge impurities\cite{Yan, Wang} or by applying a gate voltage\cite{Novoselov, Efetov}. Therefore, understanding the effect of doping on properties of graphene is of both fundamental interest and technological significance. We have performed a systematic first-principles computational study of the effect of doping on graphene's ideal strength through calculations and analyses of phonon spectra of doped graphene under external strain. Our method is based on density-functional theory (DFT) and density functional perturbation theory (DFPT) in pseudopotential plane-wave formalism, as implemented in the QUANTUM-ESPRESSO code\cite{Giannozz}. We use local density approximation (LDA) and a plane-wave cutoff energy of 150 Ry. A supercell containing a single layer of graphene and a 15 {\AA} of vacuum is used to eliminate the interaction between the graphene and its periodic images. A 30 $\times$ 30 $\times$ 1 uniform k-point mesh is used for the integration over electronic states, and a 6 $\times$ 6$ \times$ 1 q-point grid is used for the phonon calculations to obtain the dynamic matrices. The Methfessel-Paxton smearing\cite{Methfessel} with a smearing width of 0.02 Ry is used for the self-consistent calculations to ensure the convergence of the phonon frequency. The charge doping is simulated by adding/removing electrons to the graphene with a compensating uniform charge background of opposite sign.

Usually, a material's ideal (intrinsic) strength can be traced down to its bond strength. The bond strength, in principle, can be greatly affected by the amount of charge in the bond. Adding (removing) an electron to (from) the bonding (antibonding) molecular orbital of a diatomic molecule will significantly strengthen its bond, while the reverse action will weaken the bond\cite{F. Liu}. However, adding or removing electrons to a solid material (i.e. doping) has negligible effects on the overall material's strength because usually there are many more bonds than electrons. At a typical semiconductor doping level of 10$^{18}$ cm$^{-3}$, only one electron is added (or removed) among ten thousands of bonds.

Consider the case of graphene, a true 2D material of one atomic layer thick. In a graphene based field-effect transistor (FET), doping level up to 4 $\times$ 10$^{14}$ cm$^{-2}$ for both electrons and holes has been reached by electrical gating\cite{Efetov}, which corresponds to one electron being added to or removed from a few tens of bonds. Therefore, one might expect a larger effect of doping on graphene's strength than on bulk materials' strength. However, intuitively based on bond strength consideration, one would expect charge doping of either electrons or holes will weaken the graphene strength because electrons would be added (for n-doping) to the antibonding states above the Dirac point or removed (for p-doping) from the bonding states below the Dirac point, respectively. In contrast, we discover that at the typical level of doping seen in experiments, the ideal strength of graphene is substantially enhanced by either electrons or holes up to $\sim$17\%. This surprising phenomenon of pure electronic strengthening of materials is associated with the fact that the ideal strength of graphene is intricately related to the existence of Kohn anomaly and its strain enhancement, an interesting property unique to graphene.

\begin{figure*}[tbp]
\includegraphics[width=0.95\textwidth]{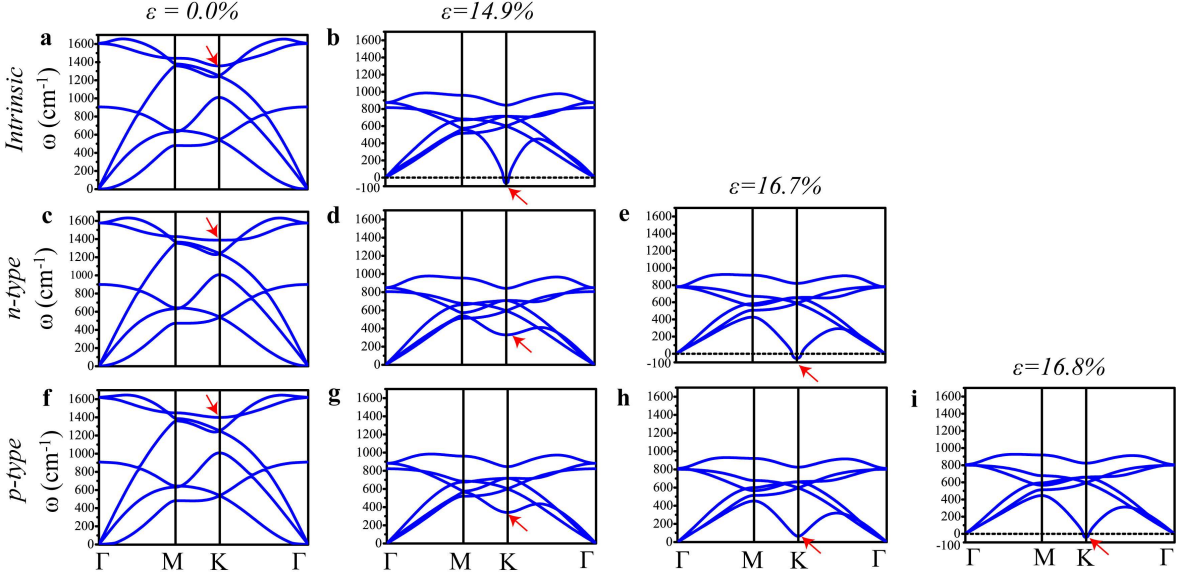}
\caption{\label{fig:fig1}  (Color online) Phonon spectra of intrinsic and doped (of 7.6 $\times$ 10$^{13}$ cm$^{-2}$ carrier density) graphene under different strains. a and b, The intrinsic graphene. c, d and e, n-type graphene. f, g, h and i, p-type graphene. The red arrows indicate the K$_{1}$ mode softened under strain.}
\end{figure*}

Figure 1 shows the calculated phonon spectra of intrinsic and doped graphene under different tensile strains, $\varepsilon$ =\emph{a}/\emph{a$_{0}$} - 1, where \emph{a} and \emph{a$_{0}$} are the strained and the equilibrium lattice constants of graphene, respectively. The primitive unit cell containing two C atoms is used for phonon calculations. Specially, we draw readers' attention to the gradual softening of K$_{1}$ mode with the increasing tensile strain, as shown from Fig. 1c to 1e and from Fig. 1f to 1i (Note that when the K$_{1}$ mode is completely softened, for convenience, we plot the imaginary part of its frequency as negative values in Figs. 1b, 1e and 1i.). For intrinsic graphene, our calculated phonon spectra without strain (Fig. 1a) agree very well with the previous calculations\cite{YanJA, prb}. The frequency of the K$_{1}$ mode becomes "negative" (the soft mode) at a critical strain of 14.9\% (Fig. 1b), signifying the ideal strength of intrinsic graphene, which also agrees well with the previous calculation\cite{Marianetti}.

Interestingly, when graphene is doped with either electrons or holes, the phonon instability under tension is greatly suppressed. Figure 1d and 1g clearly show that in the doped graphene, the frequency of the K$_{1}$ mode remains positive at the strain of 14.9\% in both n- and p-type doped graphene, indicating an enhancement of graphene strength beyond the strain of 14.9\%. At a doping level of 7.6 $\times$ 10$^{13}$ cm$^{-2}$, with the further increase of strain, the K$_{1}$ mode will eventually become soft (negative frequency) at a critical strain of 16.7\% in the n-type graphene (Fig. 1e) and of 16.8\% (Fig. 1i) in the p-type graphene, respectively. This corresponds to an enhancement of graphene's ideal strength by $\sim$13\% at this doping level.

\begin{figure}[tbp]
\includegraphics[width=0.7\textwidth]{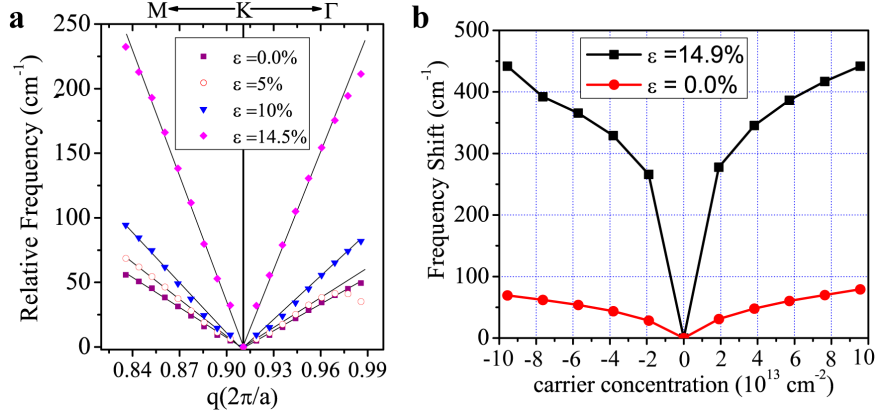}
\caption{\label{fig:fig2} (Color online) The behavior of K$_{1}$ mode under strain and doping. a, The phonon dispersions around K$_{1}$ mode under different strains: 0.0\%, 5\%, 10\% and 14.5\%. The frequencies of K$_{1}$ modes under different strain are set as the reference of zero frequency. b, The adiabatic frequency shifts of the K$_{1}$ mode (in reference to the intrinsic graphene) as a function of doping level for the unstrained graphene and a strained graphene at the 14.9\% strain, respectively.}
\end{figure}

The failure of graphene under tensile strain is known to be triggered by the mechanism of strain induced softening of zone-boundary K$_{1}$ phonon mode\cite{Marianetti}. However, the underlying reason for the softening of K$_{1}$ mode rather than other modes is not fully understood. Here, we show that this is associated with the strain enhancement of Kohn anomaly in graphene. Kohn anomaly is a sudden softening of the phonon modes occurring for certain phonons with a wavevector \textbf{q} connecting two electronic states with wavevectors (\textbf{k$^{'}$}, \textbf{k}) on the Fermi surface, \textbf{k$^{'}$} = \textbf{q} + \textbf{k}\cite{Kohn}. It arises from the screening effects of electrons on the atomic vibrations, which change abruptly the vibrations at those special \textbf{q} points. In graphene, due to its peculiar point-like Fermi surface (Dirac point), Kohn anomaly can in principle occur for phone modes at \textbf{q} = $\bf{\Gamma}$ (zone-center) and \textbf{q} = \textbf{K} (zone boundary)\cite{Piscanec}. However, the large difference in the electron-phonon coupling strength for the phonon modes at $\bf{\Gamma}$ versus \textbf{K} leads to a notable anomaly only for the K$_{1}$ mode\cite{Saha}.

Under tensile strain, the C-C bond length is stretched, resulting in the in-plane phonon modes softening. We propose that the existence of Kohn anomaly makes the K$_{1}$ mode (one of the in-plane modes) much more susceptible to softening under strain and hence responsible for graphene failure under strain. In other words, the failure is caused by a strain enhancement of Kohn anomaly that accelerates the softening of K$_{1}$ mode. The Kohn anomaly is characterized by a kink (cusp) in the phonon dispersion around the point of anomaly, i.e. the K$_{1}$ point in the present case\cite{Piscanec}. We have calculated exactly the phonon frequencies at a series of q points around K$_{1}$ as a function of strain, as shown in Fig. 2a. Clearly, all the dispersion curves are characterized by a cusp, the signature of Kohn anomaly at the K$_{1}$ point. Most interestingly, the cusp becomes deeper (i.e., larger discontinuity in the first derivative of the phonon dispersion) with the increasing tensile strain, indicating a strain enhanced Kohn anomaly, which in turn drives the rapid softening of the K$_{1}$ mode under strain.

It is known that charge doping weakens the Kohn anomaly at the K$_{1}$ point\cite{Saha}. Thus, one may expect that charge doping will counteract against strain effect and stiffen the K$_{1}$ mode under strain conditions to enhance the graphene strength. Apparently, Figure 1 already suggests that the doping enhanced graphene's ideal strength is associated with the charge stiffening of the of K$_{1}$ mode under strain (see Fig. 1d and 1g). To further reveal
this, we plot in Fig. 2b the adiabatic frequency shift of the K$_{1}$ mode as a function of doping level of electrons and holes, in the unstrained and a 14.9\% strained graphene. Clearly, we see that the frequency of the K$_{1}$ mode in both the unstrained and strained graphene increases with the increasing carrier concentration of either electrons or holes. Here, the equilibrium lattice constant of intrinsic graphene is used giving electron-hole symmetry. If the lattice constants of doped graphene are used, there will be slight electron-hole asymmetry at high doping levels. We also note that we only calculated adiabatic frequency shift at K$_{1}$ point. It has been shown that the nonadiabatic dynamic effect is important for $\Gamma$ point\cite{Lazzeri,Pisana}, which may affect K$_{1}$ point also, but we expect the effect is only quantitative without qualitatively altering our results and conclusion.

Qualitatively, doping weakens the Kohn anomaly by shifting Fermi surface away from the Dirac point, so as to move away from \textbf{k$^{'}$} = \textbf{q} + \textbf{k} condition and hence stiffening the K$_{1}$ mode. In this regard, because of the high electron-hole symmetry of graphene around the Dirac point, both electron and hole doing will shift the Fermi surface away from the Dirac point in a symmetric fashion\cite{note}, resulting in similar effects. In addition, Figure 2b shows that the doping induced frequency shift of the K$_{1}$ mode in the strained graphene is much larger than that in the unstrained graphene. The reason for such large difference is not clear, but it is consistent with the fact that Kohn anomaly has been enhanced by strain in the strained graphene as shown above (Fig. 2a), so correspondingly when the condition of Kohn anomaly is removed by doping, it may result in a larger shift of K$_{1}$ mode.

\begin{figure}[tbp]
\includegraphics[width=0.7\textwidth]{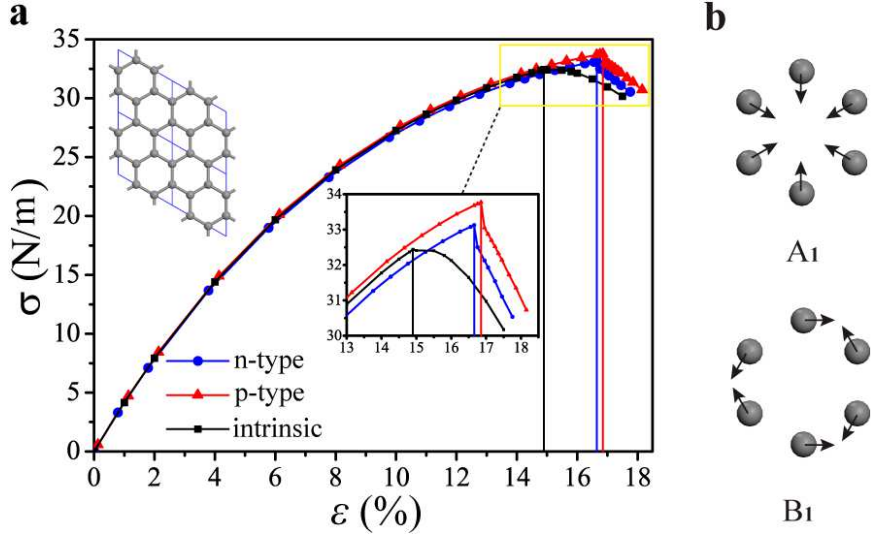}
\caption{\label{fig:fig3} (Color online) a, The stress ($\sigma$) versus biaxial strain ($\varepsilon$) for intrinsic, electron- and hole-doped (at a doping level of 7.6 $\times$ 10$^{13}$ cm$^{-2}$) graphene. The vertical lines denote the locations of structural instabilities. Inset: the graphene supercell used for calculating the stress-strain curve. b, Phonon displacements of A$_{1}$ and B$_{1}$ modes. Black balls represent the carbon atoms.}
\end{figure}

To further confirm the above mechanism of charge doping stiffened zone-boundary K$_{1}$ mode responsible for strengthening the graphene, we next determine directly the ideal strength of doped graphene by both electrons and holes, in comparison with the intrinsic graphene, by computational tensile testing. To do so, we calculate the stress tensor as a function of biaxial strain using a large unit cell of graphene with six C atoms (see the inset (I) of Fig. 3a), which is three times the size of the primitive unit cell. This choice of unit cell is made to allow the freedom of lattice distortion representative to the designated K$_{1}$ mode as done before for the intrinsic graphene\cite{Marianetti}. Group theory warrants that the vibration of K$_{1}$ mode is the combination of A$_{1}$ and B$_{1}$ modes\cite{Basko}, as illustrated in Fig. 3b, and the A$_{1}$ mode is shown to be energetically more favorable\cite{Marianetti}. Figure 3a shows the results of such computational tensile testing. By locating the breaking points (discontinuities) in the stress-strain curves, we determine the critical strains where the structural phase transition associated with the K$_{1}$ mode softening occurs. They are 14.9\%, 16.7\% and 16.9\% for the intrinsic, electron- and hole-doped graphene, respectively (see the inset (II) of Fig. 3a), which are in excellent agreement with our preceding phonon instability calculations. Above the critical strains, graphene lattice is seen to be driven to isolated hexagonal rings to start breaking down. Thus, the computational tensile testing results confirm again that the ideal strength of doped graphene is indeed enhanced by either electrons or holes through charge doping stabilized K$_{1}$ phonon mode. On a side note, the stress-strain curves of doped graphene coincide with that of intrinsic graphene below critical strains, indicating that the elastic constant of graphene is almost unaffected by doping.

\begin{figure}[tbp]
\includegraphics[width=0.6\textwidth]{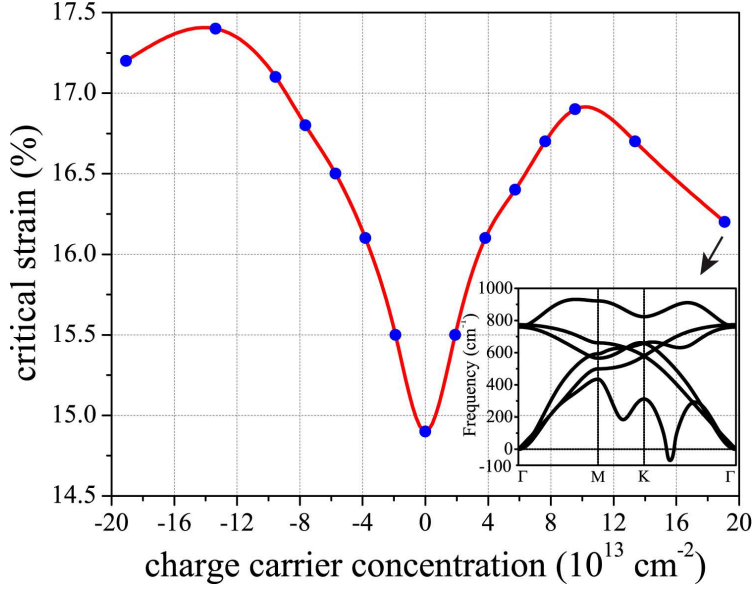}
\caption{\label{fig:fig4}(Color online) The critical strains (where phonon instability occurs) of graphene as a function of carrier concentration of electrons and holes (plotted on the negative x-axis). Inset: the phonon spectra of 1.9 $\times$ 10$^{14}$ cm$^{-2}$ electron-doped graphene under 16.2\% strain.}
\end{figure}

Finally, we determined the maximum enhancement of graphene strength can be achieved by charge doping. Figure 4 summarizes the calculation results of critical strains, where the phonon instability occurs, as a function of carrier concentration of electrons and holes (plotted on the negative x-axis). Most notably, the electronic strengthening of graphene by doping cannot go indefinitely with the increasing carrier concentration, but exhibits an upper limit. The maximum critical strain for electron doping is 16.9\% at a doping level of 9.6 $\times$ 10$^{13}$ cm$^{-2}$; that for hole doping is 17.4\% at a doping level of 1.34 $\times$ 10$^{14}$ cm$^{-2}$. This corresponds to a substantial increase of graphene strength by $\sim$13.4\% and 16.8\% for doping of electrons and holes, respectively. The slight asymmetry in the n- versus p-type graphene at high doping level is caused by the different equilibrium lattice constant of the n- versus p-type graphene. Basically, electron doping induces a lattice expansion while hole doping induces a lattice contraction, determined by the sign of "quantum electronic stress" induced by electron versus hole\cite{Hu}. Consequently, the critical strains of p-type graphene are larger than those of n-type graphene, and the difference becomes more pronounced as the doping level increases.

In fact, as the doping increases, the K$_{1}$ mode continues to stiffen because the Kohn anomaly at the K$_{1}$ point continues to be weakened. Thus, if it were only for the K$_{1}$ mode, the graphene strength would be enhanced indefinitely. However, in reality, there are many other phonon modes, and strain may induce softening of another phonon mode different from K$_{1}$ to cause failure of the heavily doped graphene. This is clearly seen in the inset of Fig. 4, which shows the phonon spectra of the 1.9$\times$ 10$^{14}$ cm$^{-2}$ electron-doped graphene under the 16.2\% strain, with a softened mode occurring away from K$_{1}$ point along the K-$\Gamma$ high symmetry line.

In conclusion, we discover a surprising pure electronic enhancement of graphene strength which is qualitatively different from the common conventional structural enhancement of material¡¯s strength. Charge doping changes the electronic structure, resulting in a phonon renormalization of graphene that improves its mechanical properties. It arises from an intriguing mechanism of charge doping counteracting against strain to affect the degree
of Kohn anomaly. It is important to note that the predicted enhancement of graphene strength up to 17\% occurs at typical doping levels easily accessible in real experimental samples, such as those used in graphene based FET devices\cite{Novoselov,Efetov}. It underscores another fascinating property of graphene with broad implications in graphene-based electromechanical devices.

We thank discussion with Li Chen. The work is support by the DOE-BES (DE-FG02-04ER46148) program. C.S. thanks Tsinghua exchange student fund for supporting her visit at U. of Utah. W. D. thanks support by the Ministry of Science and Technology of China (Grant Nos. 2011CB921901 and 2011CB606405), and the NSFC (Grant No.11074139). The calculations are done on CHPC of U. of Utah.


\begin{references}
\bibitem{note} Strictly speaking, the electron-hole symmetry will be slightly broken in the doped graphene, but it is a second-order effect without changing the overall physical picture we present.
\end{references}
\end{document}